# BACQ - Application-oriented Benchmarks for Quantum Computing

Delivering an application-oriented benchmark suite for objective multi-criteria evaluation of quantum computing performance, a key to industrial uses

Frédéric BARBARESCO
*Key Technology Domain PCC*
THALES
Velizy-Villacoublay, France
frederic.barbaresco@thalesgroup.com

Laurent RIOUX
*Research Department STI*
THALES Research & Technology
Palaiseau, France
laurent.rioux@thalesgroup.com

Christophe LABREUCHE
*Research Department STI*
THALES Research & Technology
Palaiseau, FRANCE
christophe.labreuche@thalesgroup.com

Michel NOWAK
*Research Department STI*
THALES Research & Technology
Palaiseau France
michel.nowak@thalesgroup.com

Noe OLIVIER
*Research Department STI*
THALES Research & Technology
Palaiseau, France
noe.olivier@thalesgroup.com

Damien NICOLAZIC
*Quantum Computing France*
EVIDEN
Paris, France
damien.nicolazic@eviden.com

Olivier HESS
*Quantum Computing France*
EVIDEN
Montpellier, France
olivier.hess@eviden.com

Anne-Lise GUILMIN
*Quantum Computing France*
EVIDEN
Montpellier, France
anne-lise.guilmin@eviden.com

Robert WANG
*Quantum Computing France*
EVIDEN
Paris, France
robert.wang@eviden.com

Tanguy SASSOLAS
*Calcul quantique et haute performance*
CEA LIST
Palaiseau, France
tanguy.sassolas@cea.fr

Stéphane LOUISE
*Université Paris-Saclay*
CEA, LIST
Palaiseau, France
stephane.louise@cea.fr

Kyrylo SNIZHKO
*Univ. Grenoble Alpes, CEA*
INP, IRIG, PHELIQS
Grenoble, France
orcid.org/0000-0002-7236-6779

Grégoire MISGUICH
*Université Paris-Saclay, CEA, CNRS*
Institut de physique théorique,
Gif-sur-Yvette, France
gregoire.misguich@ipht.fr

Alexia AUFFEVES
*Research Lab MajuLab*
CNRS NUS NTU UCA SU
International, Singapore
alexia.auffeves@cnrs.fr

Robert WHITNEY
*LPMMC*
CNRS
Grenoble, France
robert.whitney@grenoble.cnrs.fr

Emmanuelle VERGNAUD
*Pôle européen de compétence en simulation et HPC - TERATEC*
Bruyères le Châtel, France
emmanuelle.vergnaud@teratec.fr

Félicien SCHOPFER
*Laboratoire national de métrologie et d'essais - LNE*
Trappes, France
felicien.schopfer@lne.fr

*Abstract*—With the support of the national program on measurements, standards, and evaluation of quantum technologies MetriQs-France, a part of the French national quantum strategy, the BACQ project is dedicated to application-oriented benchmarks for quantum computing. The consortium gathering THALES, EVIDEN, an Atos business, CEA, CNRS, TERATEC, and LNE aims at establishing performance evaluation criteria of reference, meaningful for industry users.

## I. Introduction

Quantum computing promises to revolutionize multiple technical fields and activity sectors, from optimization in logistics, to simulation for research in physics or chemistry, engineering or industry, passing through cryptography. Measuring the progress towards the quantum advantage and the realization of such promises, with objectivity and reliability, is of high interest for potential end-users and crucial for the future development of the domain, now subject of hype and high competition. The challenges, especially to achieve comparable measurements, comes from the diversity of the hardware platforms, their specificities in terms of physical characteristics and applications, their maturity that can still be low, and the potential rapid evolution of the technologies.

A number of initiatives exist to benchmark the performance of quantum computers. Examples include Quantum VOLUME [1] and CLOPS [2] from IBM, SupermarQ [3] from Super-Tech or Quantum LINPACK- [4] from Berkeley Lab. The metrics used in these previous approaches are very technical and require familiarity with the technology. They therefore do not make it possible to derive operational indicators of the performance of the different families of algorithms executed on the different existing quantum computers. Dedicated to the whole value chain setting up from quantum hardware development to industrial use cases, BACQ is complementary to the benchmarking initiatives only focusing on low-level hardware physical criteria. The envisioned benchmark suite will be based on the resolution of several classes of problems covering important application domains of quantum computing which are meaningful for industrial users (see Figure 1): simulation of Quantum Physics models, Optimization, Linear System Solving and Prime Factorization. Machine Learning could be included in Optimization application domain.

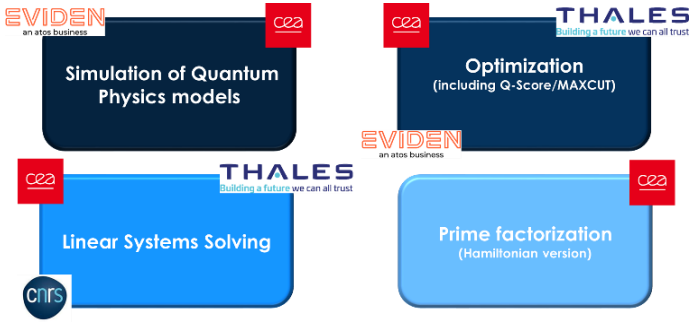

*Figure 1: Families of problems used for the QPUs benchmark*

These problems are generic and could be relevant for different branches of industries and services (chemistry, aeronautics, electronics and energy for example). Criteria will be defined for the resolution of each problem, some being hardware-agnostic and others hardware-dependent (low level): computation time, latency, problem size, approximation rate, resolution probability, accuracy, fidelity… Importantly, the project also considers energetic criteria for the evaluation of the energetic performance of the machines.

The proposed methodology consists in the aggregation of low-level technical metrics and a multi-criteria analysis via the tool MYRIAD-Q in order to provide operational performance indicators of the different quantum computing solutions and point out the service qualities of interest to the end-users. The aggregation of the criteria and multi-criteria analysis allows fully explainable and transparent notations, comparisons between different quantum machines and with classical computers, as well as identification of the practical advantages of each quantum machine with respect to specific applications. The project will address both analog machines (quantum simulators and annealers) and gate-based machines, Noisy Intermediate Scale Quantum (NISQ) and Fault Tolerant Quantum Computing (FTQC). The followed practical approach is to have a suite of benchmarks, adaptive to some extent, appropriate to the capabilities of the available machines and able to demonstrate their respective advantages, including, in the longer term, exponential speedup of specific algorithms on FTQC machines.

As part of the project, a first action has already been launched regarding Q-Score, where Eviden developed on the MAXCUT optimization problem [5], to test and validate the benchmark on different types of quantum machines.

Sharing of the benchmark suite, as broadly as possible, is an important objective to establish common reference measurement methods and ensure no bias with inclusion of all the technologies. Consultations of the technology providers, as well as end-users is key to develop an instrument that meets the needs. Once specified and developed, the benchmark suite will be available to the community for its use.

To achieve a universal tool, the project will seek to establish cooperation with similar initiatives worldwide and TQCI (Teratec Quantum Computing Initiative) seminars dedicated to "Overview of upcoming application-oriented benchmarks for quantum computing in France and abroad" will be organized on an annual basis by TERATEC, THALES and LNE. First seminar has been organized in 2023 in Thales TRT in Palaiseau [49]. Second TQCI seminar will be organized in 2024 in Reims [50].

Standardization will be another way to get consensus and adoption at large scale, considering the European committees CEN-CENELEC JTC 22 WG3 on Quantum Computing and Simulation [51] and the international ones, ISO/IEC (JTC1 WG14 on Quantum Information Technology [52] and JTC-Q on Quantum Technologies) and IEEE (P7131 "QC Benchmarking" [53], P3329 "Quantum Energy Initiative" [54] and P3120 "Architectures for QC" [55]). In France, norms are addressed by AFNOR/CN QT on Quantum Technologies [56].

At the start of the project, the machines whose access is targeted will mainly be the "Très Grand Centre de Calcul du CEA" (TGCC is an infrastructure for scientific high-performance computing, funded by GENCI) via the HQI (France Hybrid HPC Quantum Initiative, a hybrid HPC-Quantum platform and a research program [57]) action of the French national quantum strategy [6]. We will also explore all opportunities for access to machines via agreements between GENCI/CEA and EuroHPC and the QC hosting entities [7] or international organizations, or directly with QPU providers. The EuroHPC JU has signed hosting agreements with six sites across Europe to host & operate EuroHPC quantum computers. These quantum computers will allow European users to explore a variety of quantum technologies coupled to leading supercomputers.

The BACQ project, with 3-year duration and almost 4 M€ budget, including 2.5 M€ funding by France 2030 via ANR within MetriQs-France. This project precisely aims at developing, exploiting and promoting a reliable, objective and long-lasting measurement instrument for measuring the performance of quantum computing. The section II presents in more details the multiple objectives of the project. The section III provides a description of the background of each partner, their expertise in the quantum computing and benchmarking domains. In section IV, we describe the different work packages that structure this project, from the consultations and disseminations required to ensure visibility of the project, to the description of the benchmarks criteria, its tests definition and the methodology used to analyze and consolidate the results. The MYRIAD-Q tool, which is at the base of the multi-criteria analysis process and a major differentiator compared to other benchmarking initiatives, is further discussed and illustrated on a simple case in section V. Finally, section VI presents the Fast-Track initiative that has been launched regarding Q-Score and highlights the deep connections already established with a diversity of hardware providers.

## II. OBJECTIVES

### A. *MetriQs-France Program*

MetriQs-France is the French National Program on Measurement, Evaluation, and Standards for Quantum Technologies. MetriQs-France objective is to develop, exploit, and promote measurement capabilities of reference, validated and harmonized, for characterization and performance assessment of quantum technologies with reliability, impartiality and comparability: metrology, test & evaluation, international standardization are at the core of the program. MetriQs-France is coordinated by LNE, the French national metrology and testing laboratory, a public institution under the supervision of the French Ministry for Economy and Finance, in charge of industry.

### B. *BACQ project*

While looking to the future, it is critical to measure the progress towards the realization of the quantum computing promises. Application-oriented benchmarks, enabling to evaluate the real performance of quantum computing from the user perspective appear to be useful within this perspective. The challenge comes from the diversity of the hardware platforms, their specificities in terms of physical characteristics and applications, their maturity and the potential rapid evolution of the technology. Developing an objective, long-lasting, widely shared measurement instrument, to serve as a common reference.

The evaluation of the practical quantum computing performance will be considered through benchmarks close to real applications, meaningful for industrial (as well as academic) end-users. The main objective is to measure the progress towards a practical quantum advantage. In that regard, the consortium plans to perform comparisons between the different quantum computing solutions as well as to compare them with current classical computers. This benchmarking initiative will eventually highlight the assets of each quantum computing solution given specific applications.

The benchmark suite resulting from this project will be maintained by LNE, an independent and trusted third party. Thanks to their interaction with the community of end-users, this tool will be operated in order to analyze the results obtained from the machines tested with the different benchmarks using the explainable aggregation tool. LNE will establish a list of performances, maintain it over time, and update the definition of the tests. Furthermore, adoption of this initiative by French, European and eventually international partners will be encouraged thanks to the development of communication tools regarding the approach taken and visual representations of the aggregation of the results obtained by the machines from the different benchmarks. International dialogues and collaborations on the subject of quantum computers benchmarking will be promoted so that the approach, supported by MetriQs-France, is and remains an international reference. This BACQ project promotes the development of an international standardization in regard of the methods used to evaluate the specifications of quantum machines.

In view of fulfilling these objectives, BACQ has initiated collaborations and dialogue with other international benchmarking initiatives: Fraunhofer IKS & FOKUS (BenchQC Project [58]) in Germany, TNO & TU Delft (QPack Project [59]) in The Netherlands, Qilimanjaro (CUCO Project [60]) in Spain, QuIC (Use cases WG) in Europe, HamLib Project [37] (Intel, LBNL et al.) in US, Unitary Fund (Metriq project [61]) in US and QED-C [62] (QC Benchmarking WG) in US. BACQ is also involved in standardization initiatives: AFNOR National Committee on QT in France, CEN-CENELEC (JTC22 QT / WG3 Quantum Computing & Simulation) in Europe, ISO & IEC (JTC1/WG14, JTC3) at international level and IEEE (P7131 "QC Benchmarking", P3329 "Quantum Energy Initiative" and P3120 "Architectures for QC") in US.

## III. PARTNERS BACKGROUND

The consortium is composed of THALES as coordinator, EVIDEN, an Atos business, CEA, CNRS, TERATEC and LNE. Each BACQ partner brings its expertise to the consortium thanks to its previous work on quantum algorithms and quantum energetics. The main previous achievements by project partners are described in the following.

### A. *EVIDEN previous achievements*

Recently introduced by EVIDEN, the Q-score measures the real performance of quantum processors when solving optimization problems. In order to maintain uniformity of results, Q-score relies on a standard combinatorial optimization problem, the same for all evaluations (the Max-Cut problem). The score is calculated based on the maximum number of variables that a quantum technology can optimize (e.g. 23 variables = 23 Q-score or Qs). Based on open access software, Q-score is based on 3 pillars:

- **Application-centric**: Q-score is a metric based on measurements and algorithms available in the short term and on concrete operational problems;

- **Openness and ease of use**: universal and free, Q-score benefits from EVIDEN's neutral approach to quantum technologies. Does not require significant computing resources to carry out the Q-Score measurements;

- **Objectivity and reliability**: EVIDEN combines a hardware and technological agnostic approach with strong expertise acquired by working with large industrial clients and technological leaders in the quantum field.

### B. *CEA (IPhT, PHELIQS & LIST) previous achievements*

CEA IPhT has many years of experience in fundamental research on quantum information theory and on the physics of many-body quantum problems. People at CEA IPhT involved in the project have developed an open-source simulation code for qubits in interaction and in the presence of decoherence and dissipation sources (open systems and so-called Lindblad equation) using an algorithm state compression based on matrix-product-operators [8][9].

CEA PHELIQS has an extensive experience in many-body quantum problems, as well as quantum machines and devices, with in particular, participation in the QC4BW project in the German state of Baden-Württemberg which studies the possibilities of many-body quantum simulation on IBM quantum computers.

CEA LIST has know-how on the quantum annealing machines offered by D-Wave, and was able to develop a methodology for comparing classic algorithms and the use of QPU. In this research process, it was necessary to implement several error mitigation methods, some specific to D-Wave QPUs, and others more generic. Since then, the CEA LIST has increased the algorithmic footprint, on the one hand by extending the types of QPUs used (in particular QPUs with quantum gates like those from IBM), the algorithm (including the usual Simon, Grover, QAOA algorithms), and the type of problems addressed including other types of optimization problems, linear algebra as well as cryptanalysis. The roadmap therefore plans to diversify QPUs and classes of QPUs, to gain expertise on these new classes of QPU, on their use and how to improve the signal/noise ratio in the calculation results. At the same time, it will be a question of broadening the field of applications beyond the bases currently constructed.

### C. *CNRS previous achievements*

CNRS has pioneered the construction of energetic figures of merit for quantum technologies, within a tight collaboration with the CQT, Singapore. CNRS has published a position paper in PRX Quantum calling for the creation of a crossed disciplinary research line to understand and optimize the energy efficiency of quantum technologies for full-stack quantum devices [10]. A crossed disciplinary methodology has been proposed called MNR (Metric Noise Resource) and will be directly applied in the BACQ project to characterize the efficiency of machine resources. The CNRS team has already applied this methodology to minimize the energy cost of a large-scale superconducting quantum processor. Following the position paper, CNRS co-founded the Quantum Energy Initiative, an open, international and interdisciplinary community aiming to model and optimize the energy efficiency of quantum technologies. CNRS is Chair of a recently created IEEE working group dedicated to develop standards of energy efficiency for quantum technologies. International strategies of influence will benefit from the presence of CNRS in Singapore and Australia, where the QEI has found strong supports lately through the organization of two international conferences (the first conference of the QEI, Singapore, November 20-24, 2023 [63] and the first International Conference on Quantum Energy, Melbourne, December 4-6, 2023 [64]).

### D. *THALES previous achievements*

THALES worked on the study of the impact of noise on the performance of variational algorithms. This study makes it possible to determine the noise thresholds that guarantee non-degraded use of variational algorithms.

THALES has developed a multi-criteria methodology to evaluate and compare different solutions based on a set of conflicting criteria. Its main advantages are its ability to represent criteria that interact with each other, its learning modules of the model from simple information for a decision maker and its readability of the model. This approach has been used for the performance evaluation of different types of systems and in particular tracking systems. Other applications have been carried out in fields as different as trains, space, cyber security and naval defense. In order to obtain an evaluation model shared by several users, collegial interview sessions can be organized. The preference information provided is typically preference intensities between different options. The Thales Research & Technology tool takes into account a plurality of opinions by offering responses in the form of an interval. This makes it possible to take into account similar but not identical opinions.

## IV. WORK PACKAGES DESCRIPTION

The goal of this section is to present the work planned during the project for each work package. The project is organized around six work packages as illustrated in the following Figure 2.

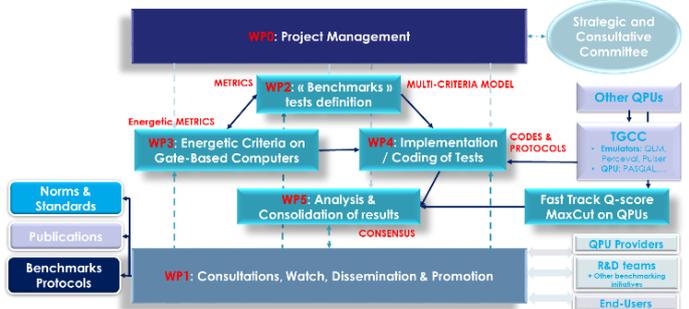

*Figure 2: Organization of work packages*

The set of work packages is composed of two non-technical ones. WP0 corresponds to the "study management" and defines the management of the project on a daily basis as well as interactions with the working groups in France in order to ensure the consistency of the work with national ambitions and strategy. WP1 focuses on consultations, monitoring, dissemination and influence. Its objectives are to ensure the visibility of the project at the national and international level, to promote its results and to share its ambitions with other studies underway at the global level. Part of this work package will be devoted to the development of the first standards of energy efficiency for various quantum computing paradigms at IEEE, within the framework of the QEI P3329 IEEE Working Group [11].

The remaining four technical work packages aim to define and implement the tests, collect the results and analyze them. In WP2, the goal is to identify and construct relevant criteria which will be used to set up the benchmarks. WP3 focuses on the study of energy efficiencies regarding some of the algorithms considered in this project. The work package will study in detail the case of a gate-based quantum computer executing a simple algorithm such as the resolution of a linear system. The tasks proposed in WP4 are the core of the project insofar as the objective is to implement and develop the test protocols proposed in WP2 on real quantum computers (and possibly emulators). Finally, WP5 consists in the analysis and consolidation of results. It aims to collect the codes and the first experimental results previously achieved, and carry out a critical and statistical analysis of the results. The results will be consolidated using successive refinements with feedbacks from QPU providers on how to better use their machines.

### A. *WP1 Consultation and dissemination*

The aim of this work package is to ensure that the project is well positioned in a very dynamic scientific and international context. The main tasks are monitoring, liaising with stakeholders, making the project more visible, disseminating, promoting and exploiting the results.

In this way, we will be able to monitor the technologies studied and the benchmarking initiatives at global level, to consult industrial users and suppliers within the framework of the Working Groups (WG), to establish and maintain an international dialogue in order to share ambitions with other studies underway, as well as promote the results in standardisation works.

First, a review of the state-of-the-art benchmarking methods and a scientific watch are essential.

Consultations with industrial stakeholders is necessary to develop and maintain relations with the various international actors, especially technology suppliers and users. Stakeholder consultation helps to ensure that the benchmarks developed meet the needs of technology suppliers and industrial users. Regarding technology suppliers, the dialogue ensures in particular that the benchmarks are adapted to current machines. It is also important to facilitate access to machines for the needs/requirements of the project. Liaisons will be established with suppliers of all the technologies (solid state, atoms, ions, photons…). The contacts made for the Q-score FastTrack action give a good example of the objective. Regarding the industrial users, the widest possible range of sectors of use to be covered, in conjunction with the definition of operational indicators relating to specific quality of services (via the use of MYRIAD-Q).

Collaborations with international benchmarking initiatives are necessary with the objective to develop benchmarks that will be widely adopted and serve as a common reference. These collaborations with similar initiatives at the European level and worldwide is essential. The complexity and width of the benchmark subject make it also necessary to consider complementary actions. Collaborations are envisioned and launched with R&D teams, but also consortiums and more structured and coordinated actions.

Standardization is another important aspect of this work package. The goal is to promote and enhance the benchmarks developed, making them international standards for measuring the performance of quantum computers. It is expected that the evaluation results, as technical specifications, will also feed the development of a wide range of standards. The impact in this field will be based on the participation of partners in the main Standard Development Organizations addressing quantum technologies. At the national level, this is the French national AFNOR standardization committee on Quantum Technologies. At the European level, this is the CEN-CENELEC Joint Technical Committee JTC22 on Quantum Technologies. At the international level, this is mainly the recently established ISO/IEC Joint Technical Committee JTC3 on Quantum Technologies, which will take over the ISO/IEC JTC1 WG14 on Quantum Information Technology. There is currently a project dedicated to application-oriented benchmarks under development at CEN-CENELEC JTC22. Partners are also taking part to IEEE Standards Association projects related to benchmarks like P7131 "Quantum Computing Performance Metrics & Performance Benchmarking" or P3329 "Quantum Computing Energy Efficiency".

Finally, dissemination and communication activities are particularly important to identify communication channels, forthcoming conferences, workshop, to publish the project results in high-impact peer-reviewed journals and to communicate broadly on the progress of the project as a whole.

The first TQCI seminar has been organized by TERATEC, THALES and LNE at Thales Research & Technology in May 2023 with about 100 attendees of R&D teams, Quantum Processing Unit (QPU) providers, end-users, policy makers, governmental agencies from Europe, US, Japan, and Singapore [49].

B. *WP2 Benchmark tests definition*

Based on the state of the art on the different benchmarks, the objective of this work package is to identify the different criteria and the associated algorithmic approaches. To do this, the partners will help identify the criteria and divide them into different classes (criteria linked to the problem addressed by a benchmark, criteria linked to the method or different calculation methods and criteria linked to the computers). Finally, work package identifies the generic multi-criteria approach allowing the construction of an aggregate score.

As seen in the introduction, the set of benchmarks is divided into four categories: Optimization problems (including, but not limited to, EVIDEN's MaxCut QScore), linear system solving, quantum physics simulations and prime factorization. To this end, the set of problems constituting each benchmark family should be viewed as representative enough of the field (although pursuing exhaustivity would be out of the scope of the project at this time).

*1) Base principles of benchmark design*

In order to design useful and relevant benchmark programs, it is necessary to adhere to a set of important principles.

As current quantum machines have a relatively high level of noise, it is of uttermost importance to compare the measurements we produce against a pure random machine. As these QPUs are good random generators, the usefulness of quantum computing that would be undistinguished from random results must be considered as null.

Every family of benchmark should provide at least two different applications per family whenever it makes sense (i.e. except for Prime Factorization which describe one and only one possible application, even though the implementation can still vary),

As the strong points of a given quantum processor can differ from another, the exact algorithm would better be fitted to match these strong points for a given application. Nonetheless, the design and specification process must be accurately and thoroughly followed and described so that it will be easier to create versions adapted to new hardware platforms. One simple illustration of that can be illustrated by the fact that quantum annealers (e.g. D-Wave QPUs) cannot apply the same algorithm as gate-based QPUs.

Last but not least, each design processes toward the implementation in software of each benchmark program must be thoroughly detailed so that it must be easy for a third party to re-implement the benchmark program from its documentation but also to design easily a new implementation for a new QPU.

The following sections describe each of the four benchmark families.

*2) Optimization*

Optimization is a field for which quantum computing is believed to be the most apt at for short to mid-term industrial applications. Nonetheless, it is also believed that BQP, the set of problems solvable in polynomial time on a quantum computer is different from NP, therefore one should not have unrealistic expectations from quantum computing approaches. However, it is a domain of application for quantum computing of the highest importance, because it may improve the time to solution, or the quality of approximate solution, or, even more importantly, the energy required to achieve a given level of quality for a solution.

The family of optimization problems and programs will be articulated around 3 to 4 sets of problems:

- MaxCut (as part of Eviden's QScore), as an example of NP-hard quadratic binary unconstrained model [5],

- Maximum Cardinality Matching, as an example of Polynomial constrained problem which is famously hard for annealing approaches [13],

- Higher Order Binary Optimization (HOBO) problems, which are not necessarily NP-hard, but will probe the case of non-quadratic problems which can be important in the general case, e.g [14],

- Optionally, some NP-hard constrained problems.

The Maximum Cardinality Matching family of problems is known to be only polynomial but they laid the base to demonstrate the worst-case convergence of Simulated Annealing [15]. CEA-LIST was pioneer in utilizing the $G_n$ series of problem to probe the capabilities of D-Wave Quantum Annealers [13]. This work has since been extended by other research teams in the world [16]. Work is currently underway to extend the study to the latest iterations of D-Wave QPUs and expose the best strategies toward obtaining the best results. These results should be easily extended to other Quantum Analog QPUs (e.g. Pasqal) and QAOA approaches to solve optimization problems.

HOBO problems are important because, while it is always possible to convert them to quadratic problems by introducing ancilla variables (a process called quadratization [17]), doing so introduces many complications for the transformed problems. Therefore, QPUs that allow for avoiding such transformations would take an increased edge against QPUs that would necessitate a quadratization of the problem before solving.

*3) Linear Systems*

Linear algebra in general provides a set of problem that are as important as optimization problems in term of applications, and is a field where ideal quantum computing has a known exponential speedup [18]. Although it is clear that no short term exponential computing advantage can be expected, it remains an important field to investigate.

To that end, two sets of applications stand out:

- Linear system solving, which is the set for which HHL will be the most relevant algorithm for gate-based QPUs.

- Eigenvalues – eigenvectors of a given matrix, as it is a family of application in which some hybrid algorithms like VQE [19] could be applied.

For the latest, VQE is already known as a family of algorithms which can be a potential short term application of quantum computing.

*4) Quantum physics simulations*

Quantum physics problems arise in various fields of science. For instance, in solid-state physics the understanding of new quantum phenomena and the design of new materials with potential applications often require to simulate some quantum problems with many electrons. Quantum physics with many electrons also play a central role in chemistry. They are also central problems in other domains of basic research, such as in particle or nuclear physics an even quantum information theory. Among the simple but yet nontrivial many-body Hamiltonians defined on lattices one can mention quantum spin-1/2 models (Heisenberg interactions, Ising model in transverse field, …) spinless fermion models, spin-1/2 fermion models (Hubbard model) or boson models (ex: Bose-Hubbard). Simulating these quantum many-body systems is notoriously difficult on a classical computer, for the same reason as a quantum computer is more powerful than a classical one, namely the Hilbert space dimension of a quantum system grows exponentially with the number of particles. However, with an analogue quantum simulator or with a digital (gate based) quantum computer, one would in principle be able to study many-body systems which are untractable on a classical machine. In that context one can mention two distinct tasks: 1) from a given Hamiltonian $H$, determine the expectation values of some observables (including $H$ itself) in the ground-state of $H$. 2) Given a simple (product) initial state $|\psi_0\rangle$, determine the expectation values of some observables in the time-evolved state $|\psi_t\rangle = \exp(-iHt/\hbar) |\psi_0\rangle$. The ground-state problem (1) can for instance be tackled using VQE on a gate-based machine or via a (quasi) adiabatic evolution on an analogue simulator. The dynamical (also called quantum quench [20, 21] problem (2) can be studied via a discretization of the time evolution (Trotterization) on gate-based architectures, and somehow more directly with the native Hamiltonian dynamics on an analogue device (provided the interactions between the qubits can be tuned to match the Hamiltonian $H$ of interest).

*a. Analogue simulators*

Various types of benchmarks have been proposed to assess the performance of gate-based machines but there are much fewer proposals for analogue quantum simulators. On an analogue machine the precision of the outcome depends on the capacity to accurately i) prepare an initial state ii) control the (possibly time-dependent) Hamiltonian and iii) perform measurements on the final state. In this study we will construct and test protocols based on some exactly solvable spin model Hamiltonians $H$. This will ensure that the exact final state expectation values are known (can be computed), even for a large number of qubits (scalability). Examples include models where spin-1/2 are arranged one-dimensionally (spin chains), such as the XY chain or the transverse-field Ising chain (both exactly soluble by mapping to free fermions), or the XXZ chain (solvable by Bethe's Ansatz). In the class of systems which can be simulated efficiently but not solved exactly, we can also add one-dimensional problems with a gap in their spectrum of excitations, which can be simulated by the Density Matrix Renormalization Group (DMRG) [22][23].Even if these models are well understood and exactly soluble, their ground states are not trivial states, but correlated and entangled. In that sense, they are representative of the states which can arise in realistic quantum physics problems.

Due to the presence of noise the state realized experimentally will be a mixed state $\rho$. Measuring experimentally the fidelity $F = \langle\psi_t|\rho|\psi_t\rangle$ between $\rho$ and the theoretical target state $|\psi_t\rangle$ is a priori not an easy task if one is looking for a scalable protocol which can pushed to a large number of qubits. One option is to use the direct fidelity estimation method of [45]. In parallel, we may also choose a set observables $\hat{O}_{\alpha=1\cdots n}$ and quantify how the measured expectation values $m_\alpha = \text{Tr}[\rho\hat{O}_\alpha]$ differ from the ideal values $o_\alpha = \langle\psi_t|\hat{O}_\alpha|\psi_t\rangle$. This would lead to a proxy for the infidelity: $G = \sum_\alpha |m_\alpha - o_\alpha|$. A natural choice for the set observables would be the one-qubit observable $X_i, Y_i, Z_i$ and the 2-qubits observables $X_iX_j, X_iY_j, X_iZ_j, Y_iY_j, \ldots$ ($X_i, Y_i$ and $Z_i$ are the Pauli operators acting on qubit $i$). This way, one would be able to quantify the errors on physically relevant quantities (local expectation values and 2-point correlations).

*b. Gate-based devices*

For gate-based machines, we will simulate the time evolution of a BCS model on a quantum computer. The choice is dictated by several considerations: (i) being originally fermionic, the model has a representation in terms of spins – that is, it can be directly mapped to qubits; (ii) the model requires all-to-all interactions, which enables assessing the qubit connectivity; (iii) recently a methodology enabling one to assess the significance of two-qubit-gate errors and crosstalk errors based on the BCS model simulation has been developed [49]; (iv) the model is integrable, which, in principle, enables verification of large-size simulations, when these become practical.

*5) Prime Factorization*

The discovery of Shor's algorithm [24] for prime factorization is one of the results which generated immense expectations on the part of quantum computing. While it is clearly unrealistic to expect that this particular algorithm can be put to applications without having first developed fault-tolerant quantum computers, there exists other possible approaches. One is utilizing Hamiltonian based approaches [25]. The basic principle is rather simple because it simply requires the minimization of the following cost function:

$$C(p_1, p_2) = (N - p_1p_2)^2$$

Where $p_1$ and $p_2$ are 2 possible prime numbers that ought to be found and $N$ is the integer which is the subject of the prime factorization.

Approaches that are not based on Shor's algorithm are not expected to provide any kind of computational advantage, it can however not be ruled out at the moment that it can provide some kind of advantage on the energy side.

C. *WP3 Energy criterion by resource estimation for the resolution of simple algorithms on gate-based computers*

For any given task, an energy efficiency can be parametrically defined as the metric chosen to quantify the performance of the task, divided by its energy cost [10, 44, 48]. The efficiency is thus tightly dependent on the choice of metric, as well as the list of energy costs.

The objective of the work package is to propose and study the behavior of such energy efficiencies for some of the useful algorithms studied in this project. In principle, all machines can give rise to such figures of merit. Our approach will rely on the MNR methodology introduced in [44], which consists in finding relations between Noise, Metric of performance and Resource cost, then exploit these relations to minimize the resource cost under the constraint of a given performance.

In the spirit of the BACQ project, we will care to remain as hardware-agnostic as possible to propose useful and flexible numerical tools to end-users and industrials. The focus shall thus be put on algorithmic resources, the connection with energy costs being handled via effective models. In close connection with the HQI project of the French quantum strategy, we will build a user-friendly interface accessible to any hardware provider can use to estimate the energy efficiency of the quantum processor at play.

D. *WP4 Implementation of the first tests*

This set of tasks is at the heart of the project insofar as the objective here is to implement (code) and develop on real quantum computers (possibly in parallel with the use of emulators) the protocols of tests which will have been proposed on all the tasks of WP 2.

After designing, within WP2 and WP3, a set of problems and protocols to be used to serve as the basis of the benchmarks, the WP4 will tackle the tasks of implementing these protocols and testing them on emulators as well as on real quantum devices. This part of the project will also presumably imply adapting and fine-tuning the problems and protocols constructed in WP2 and WP3 to ensure that the benchmarks produce meaningful results when executed on emulators (including noise whenever possible) and/or confronted with the physics of real quantum devices. These studies will naturally take advantage of the access to quantum machines made possible thanks to the HQI action, as well as of the (classical and quantum) computing infrastructures of the TGCC.

The codes developed here, as well as the associated documentation, will as much as possible be made publicly available (repository system) so that the quantum computing community – including QPU providers – will be able to use them and contribute to their development by proposing improvements. In terms of development tools (languages, libraries, version control …) the priority will naturally be given to widely adopted, portable and open solutions.

Our collaboration with GENCI has established a protocol for accessing the advanced infrastructure at the TGCC. This development is a significant milestone for the BACQ project members, enabling access to a range of quantum emulators, including Qaptiva (previously known as QLM), PERCEVAL, and PULSER.

Looking forward, early 2024 is anticipated to be a transformative period for our project. At this time, BACQ members are expected to gain access to a Quantum processor developed by PASQAL, boasting a capacity of 100 to 200 Qubits. This will significantly enhance our computational capabilities.

Moreover, the timeline extends to late 2024 or more likely 2025, when access will be granted to another Quantum processor by QUANDELA, this one equipped with up to 10 qubits.

E. *WP5 Analysis and Consolidation of results*

This task collects codes and first experimental results carried out in WP4. This requires critical and statistical analysis of the results and successive refinement with feedback from QPU providers on how to better use QPUs.

WP5 delves into the evaluation and synthesis of the experimental outcomes achieved in WP4. This entails compiling and consolidating the codes and experimental data, conducting a statistical assessment of the results, and refining the findings with input from QPU providers. Thales spearheads the task of orchestrating the collection and consolidation of the acquired results and establishing a results hosting platform. CEA-List will handle the utilization of the Thales Myriad environment to aggregate the outcomes, analyze the performance of various QPUs, and generate a synthetic set of values to contrast the various hardware platforms. This comprehensive analysis will yield a nuanced comprehension of QPU performance and orient future development endeavors.

## V. MYRIAD-Q

A. *Basic concepts*

We are given a set $N = \{1, ..., n\}$ of metrics represented by spaces $X_1, ..., X_n$ respectively. The quantum solutions we consider are characterized by a value over each metric, and is thus considered as an element in $X = X_1 \times \cdots \times X_n$. The aim of Multi-Criteria Decision Aiding (MCDA) is to determine how the element in $X$ shall be compared, formalized by a so-called preference relation $\succcurlyeq$ over $X$. For two alternatives $x, y \in X$, "$x \succcurlyeq y$" means that $x$ is at least as preferred as $y$. It is important to defining such binary relation $\succcurlyeq$ necessarily to incorporate some expert(s) or decision maker(s) preferences. The comparison of options clearly depends on which metrics are the most important ones, what is the benefit of improving the value of a metric of one unit, and so one. MCDA aims thus at capturing the preferences of DMs (decision makers).

We look for a numerical representation $u: X \to S$ of the preference with $S \subseteq R$ being a scale, such that $x \succcurlyeq y$ if and only if $u(x) \geq u(y)$ [42]. The construction of $u$ requires two different operations, namely to normalize the metrics as they are given in different units (e.g. a time in seconds vs. a consumption in Watt) and to aggregate the different metrics to come up with a single score. These two operations are then performed separately in the so-called decomposable model [26]: for $x = (x_1, ..., x_n)$, we write

$$u(x) = F(u_1(x_1), ..., u_n(x_n)),$$

where $u_i: X_i \to S$ is the *marginal utility function* on metric $i$, and $F: S^n \to S$ is an *aggregation function*. We call *criterion* the utility function $u_i$ attached to a metric.

As the number of metrics is usually significant ($\gg 2$), the outputs of criteria are aggregated through a hierarchy. The idea is to introduce several nested aggregation functions. Each intermediate aggregation node introduces an intermediate score, which has a relevant meaning for the user and eases the elicitation phase. Figure 3 shows a possible organization of criteria. At first level, we find the different problem that we assess and then the list of metrics for each problem.

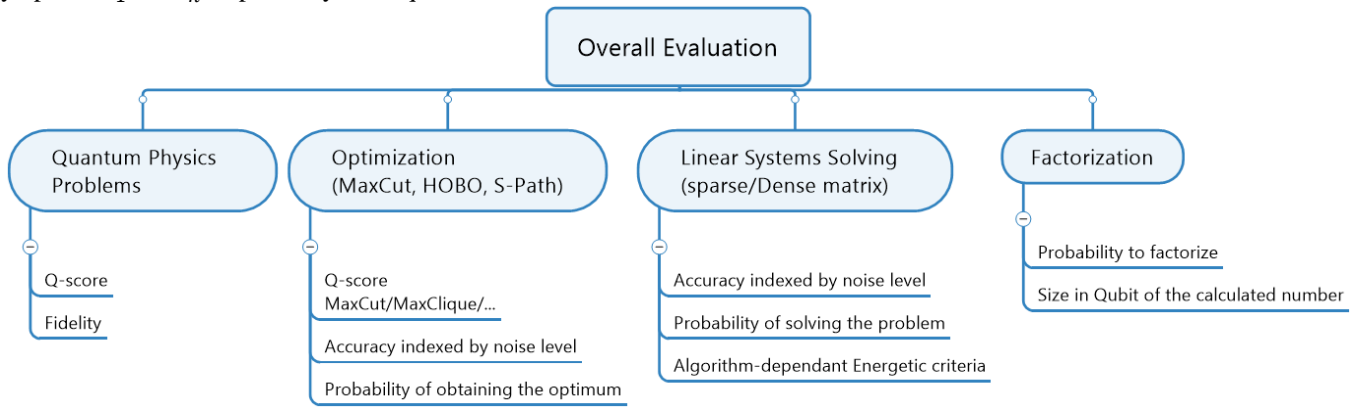

*Figure 3: Sketch of a hierarchy of criteria*

Scale $S$ is usually chosen to be bounded – typically $S = [0,1]$ meaning that there is a value of the metric above which (under which respectively) the criteria perfectly (or almost perfectly) satisfactory (not satisfied at all, respectively). Yet, the aim of MYRIAD-Q is not to only assess the current quantum technologies at present time, but to propose an assessment module that will be used for several years. However, we cannot forecast today the levels that will be reached in several years. Using a bounded scale cannot allows us to use the same model across several years. We propose thus to use the unipolar scale $S = [0, +\infty)$ for which there is a lower bound but no a priori upper bound on the performance.

The main advantages of MYRIAD-Q Multi-Criteria Approach is illustrated in Figure 4.

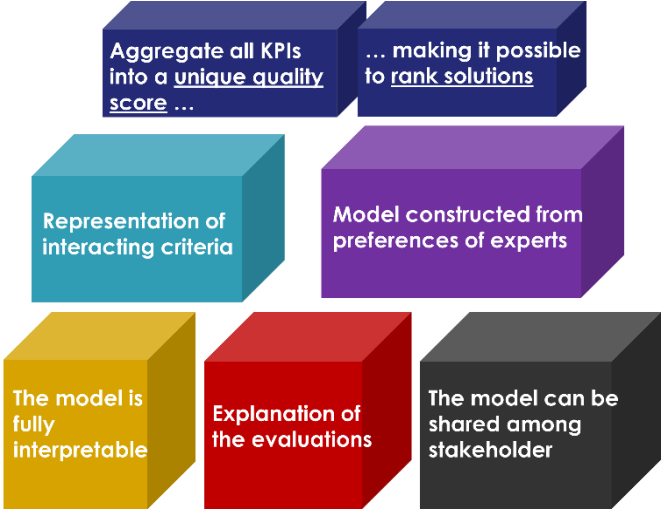

*Figure 4: MYRIAD-Q Main Advantages QPU End-Users*

### B. Construction of the model

The construction of the MCDA model is organized around three steps.

#### 1) 1st step: structuring phase

Given the plurality of QPUs, we first need to identify the subsets of QPUs that make sense to be compared. A MYRIAD-Q model is then constructed for each subset. For instance, it might not be relevant to compare Quantum Annealers with NISQ.

Once we have selected the scope of a MYRIAD-Q model, the multi-criteria approach focuses firstly on identifying operational metrics which have a real operational impact (fundamental objectives for experts) and not technical metrics which only have an indirect impact. Then the associated criteria are organized hierarchically.

#### 2) 2nd phase: normalization

It does not make sense to directly combine the metrics as they are given in different units. The aim of the marginal utility functions is to normalize the metrics and transform them in the same scale. This normalization corresponds to an *interval scale* in the terminology of measurement theory. An interval scale is given up to an affine transformation. One thus needs to define two reference levels to entirely fix it. For scale $S = [0, +\infty)$, we introduce the following two reference elements: $B_i$ (it stands for Bad) is an element of metric $X_i$ that is judged not satisfactory at all, and $G_i$ (Good) is an element of metric $X_i$ that is judged *satisficing*. This latter level is a central ingredient of the Bounded Rationality of Simon. It is a contraction of satisfying and sufficient. The bad element is the saturation level from below (no element has a worst utility); and there exists elements that are preferred to the satisficing level.

In order to ensure commensurateness among criteria, we impose that all reference levels of the same nature across the criteria are assigned to the same utility:

$$u_1(B_1) = \cdots = u_n(B_n) = 0$$

$$u_1(G_1) = \cdots = u_n(G_n) = 1.$$

Function $u_i$ shall be specified for any value of the metric. In our application, the metrics are continuous (e.g. a consumption in W) or can be assimilated to a continuous variable (e.g. Q-score returning an integer). For a continuous attribute, function $u_i$ can take different shapes. A concave shape indicates that an increase of one unit in the metric space has more impact on the utility for small values of the metric than for large values. The opposite behavior holds for convex utility functions. From psychology and cognitive science, marginal utility functions representing human behaviors are concave -- see e.g. Saint Petersburg paradox [27], risk aversion in decision under uncertainty [28], fairness in social science [29].

To fully specify the utility function, we identify in $X_i$ a finite number of elements $\widehat{X}_i = \{x_i^1, \ldots, x_i^{p_i}\}$ including the two reference elements $B_i$ and $G_i$. The DM is first asked to rank order these elements from the worst to the best one. Then, given two elements $x_i^k$ and $x_i^l$ of $\widehat{X}_i$ for which $x_i^k$ is strictly preferred to $x_i^l$, the DM is asked to identify the gain in satisfaction (attractiveness) by going from $x_i^l$ to $x_i^k$. The response should be given in the following finite scale: "Very weak", "Weak", "Moderate", "Strong", "Very strong" and "Extreme" [34]. The values of the utility for the elements in $\widehat{X}_i$ are derived from these answers. The values of the utility for the others elements of the metric are obtained by interpolation.

#### 3) 3rd phase: aggregation

As already mentioned, the aggregation function $F$ is organized hierarchically. The first phase returns a tree described by a set $M$ of nodes including a root $r \in M$ and by a function $P: M \to 2^M$ returning the set of parents $P(k)$ for every node $k \in M$. The leaves of the tree are the set $N$ of criteria. An aggregation function $F_k: S^{P(k)} \to S$ is defined at each aggregation node $k \in M \setminus N$.

One needs to elicit each aggregation function $F_k$. The simplest aggregation function is the weighted sum $F_k(a_{P(k)}) = \sum_{l \in P(k)} w_l a_l$, where $w_l$ is the weight of node $l$. This model cannot capture real-life decision strategies as it assumes that all its inputs are independent.

In order to overcome the limitations of this model, we propose to use the Choquet integral [30], which has the capacity to represent interactions among criteria [31]. Its so-called 2-additive version is a good compromise between simplicity and expressivity of the model [31]. It takes the form

$$F_k(a_{P(k)}) = \sum_{l \in P(k)} w_l a_l + \sum_{\{l,m\} \subseteq P(k)} \left( w_{l,m}^\wedge \min(a_l, a_m) + w_{l,m}^\vee \max(a_l, a_m) \right)$$

where coefficients $w_l$, $w_{l,m}^\wedge, w_{l,m}^\vee$ are non-negative coefficients that sum-up to one. Compared to the weighted sum, the min terms represent *complementarity* between the two criteria $l, m$ (necessarily both scores have to be high to produce a high overall score; in other words, if one score is larger than the other one – say $a_l > a_m$ – then the better evaluation on criterion $l$ is penalized by the worst score on criterion $m$). Likewise, the max terms represent redundancy between the two criteria $l, m$ (it is sufficient that one score is high; in other words, if one score is larger than the other one – say $a_l > a_m$ – then the bad evaluation of criterion $m$ is saved to a degree $w_{l,m}^{\vee}$ by the better evaluation on criterion $l$.

In order to elicit the aggregation function $F_k$ at node $k$, we adopt the approach proposed in [30]:

- Alternative $B_{C(k)}$ which is Bad on all inputs;
- Alternative $(G_i, B_{C(k)\setminus i})$ which is Good on input $i$ and Bad on all other inputs;
- Alternative $(G_{\{i,j\}}, B_{C(k)\setminus\{i,j\}})$ which is Good on inputs $i$ and $j$, and Bad on all other inputs.

The DM is asked to rank order these fictitious alternatives from the worst one to the best one. Then this shall be refined by providing an intensity of preference in the finite scale "Very weak", "Weak", "Moderate", "Strong", "Very strong" and "Extreme". The parameters $w_l$, $w_{l,m}^{\wedge}$ and $w_{l,m}^{\vee}$ are deduced by an algorithm [26].

C. *Illustration*

For illustration purpose, we consider the aggregation of two metrics: Q-score MaxCut and Q-score MaxClique. Figure 5 shows the multi-criteria tree. The two leaves of the tree (with a "$U$" on the left – standing for Universe) are the two KPIs, the two nodes just above them are the normalisation of the KPIs (with a "$C$" on the left – standing for Criterion), and the top node is the overall score (with a "$A$" on the left – standing for Aggregation).

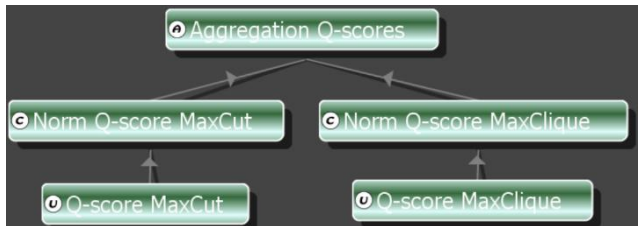

*Figure 5: Tree based on the two KPIs*

The next sections shows how the criteria and aggregation are constructed.

*1) Construction of the utility functions*

Consider attribute "Q-score MaxCut". We start by identifying the two reference elements on the metric space. The first reference element (denoted by $B_1$), which is judged not satisfactory at all, corresponds to the worst value, which is clearly 0. The satisfying element $G_1$ is taken as the largest value of the metric that is reached at current time. For the Max-Cut problem, the largest Q-score is roughly 1000 on D-Wave Classical Solver (Simulated Annealing) [50]. The choice of 1000 is based on the current number of qubits available on 'high-end' QPUs. This Q-score value is highly optimistic for today's standards and undoubtedly a challenge for the next few years. It's an arbitrary reference point.

To set the utility function for intermediate Q-score values, we consider three values obtained for different computers:

- standard PC with CPU 4 cores/8 threads @3.7 GHz: Q-score=17. This score shall not be considered as a reference. It is given to fix an order of magnitude. It is incorporate such value in the normalization scale;
- D-Wave 2000Q: Q-score=70 [32][33]
- D-Wave Advantage: Q-score=140 [32][33]

In order to fix the utility of these three intermediate Q-scores, we ask an expert to fill the intensity of preferences between elements of the five values. As an example, we set as illustration: 0, 17, 70, 140 and 1000. According to Figure 6, 17 is weakly (intensity 2) preferred to 0, 70 is strongly (intensity 4) preferred to 17, 140 is strongly (intensity 4) preferred to 70, and 1000 is very strongly (intensity 5) preferred to 140.

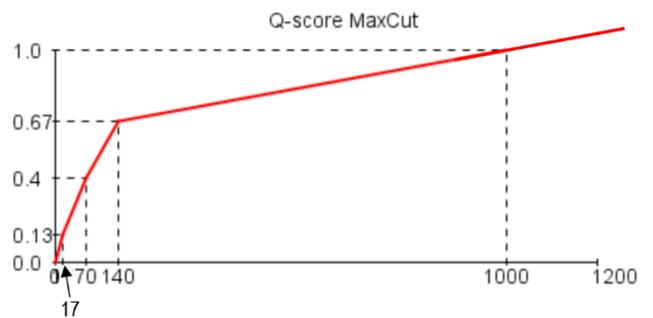

*Figure 6: Elicitation of the intensity of preference among the 5 values on "Q-score MaxCut"*

This yields the utilities shown in Figure 7

| Universe | Utility |
|---|---|
| 0.0 | 0.0 |
| 17.0 | 0.133 |
| 70.0 | 0.4 |
| 140.0 | 0.667 |
| 1000.0 | 1.0 |

*Figure 7: value of the utilities for the fives elements of "Q-score MaxCut"*

The utility function is then obtained by interpolation – see Figure 8.

*Figure 8: Utility function of Q-score MaxCut*

A similar approach is applied to KPI "Q-score MaxClique". Proceeding in the same manner, the two reference elements are $B_2 = 0$ and $G_2 = 1000$, the intermediate elements are 12 for PC with CPU 4 cores/8 threads @3.7 (here also not considered as a reference), 70 for D-Wave 2000Q and 110 for D-Wave Advantage. We obtain the utility function depicted in Figure 9.

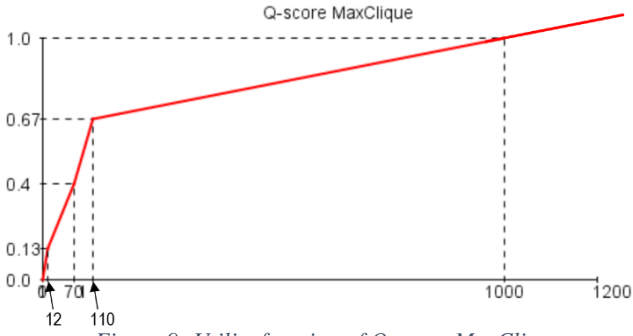

*Figure 9: Utility function of Q-score MaxClique*

*2) Construction of the aggregation function*

The aggregation functions aggregates the normalized scores of "Q-score MaxCut" and "Q-score MaxClique". We introduce the following four fictitious alternatives:

- $\langle B_1 B_2 \rangle$: This alternative takes bad levels on the two criteria and has a Q-score of 0 on both problems. This is the worst possible alternative;

- $\langle G_1 B_2 \rangle$ : This alternative is satisfycing on MaxCut problem and very bad on MaxClique;

- $\langle B_1 G_2 \rangle$: This alternative is satisfycing on MaxClique and very bad on MaxCut;

- $\langle G_1 G_2 \rangle$: This alternative is satisfycing on both problems. This is the best alternative among the four.

On needs first to compare these four alternatives. We only to compare the two intermediate alternatives. MaxCut and MaxClique optimization problems belong to the same class of NP-hard problems. However, MaxCut is considered as among the hardest problem to solve in this category. As a result we prefer option $\langle G_1 B_2 \rangle$ to $\langle B_1 G_2 \rangle$. The intensities of preference are given in Figure 10. In general, when a QPU is good at a problem, it is also good at the other one. Hence, there is no so much difference between $\langle G_1 B_2 \rangle$ and $\langle G_1 G_2 \rangle$. These preferences shall be considered as examples of parameterization of the MYRIAD tool. Of course, one may enter different preferences.

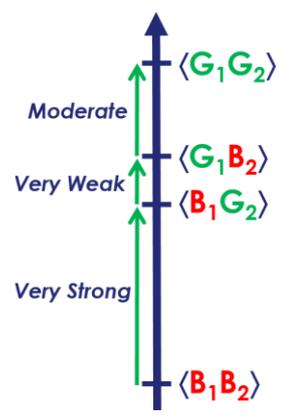

*Figure 10: intensity of preferences for the aggregation*

The parameters of the aggregation are shown in Figure 11. The upper gauges show the relative importance of the two criteria. Without surprise, the "Q-score MaxCut" is slightly more important than "Q-score MaxClique". The lower gauge represent the interaction level between the two criteria. It is negative here, indicating redundancy between the two criteria.

This is explained as when a QPU is good at solving MaxCut, it is also good at solving MaxClique, and vice versa.

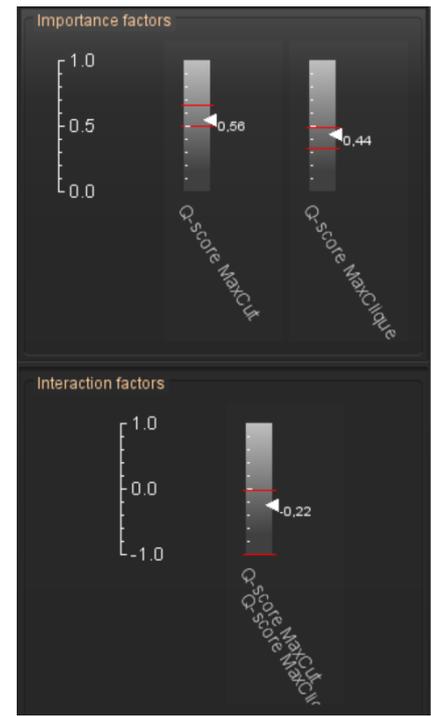

*Figure 11: Interpretation of the aggregation model*

The aggregation model is graphically represented through the pie chart of Figure 12. We see that there are three pieces – each of them representing a term in the expression of the aggregation function. The two pieces in red correspond to a criterion alone (one for "Q-score MaxCut", and one for "Q-score MaxClique"). The last piece (in green) corresponds to redundancy between the two criteria – meaning that this piece is represented to the level of the best score among "Q-score MaxCut", and one for "Q-score MaxClique" (Figure 12).

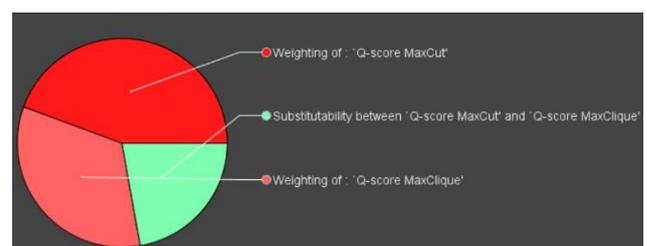

*Figure 12: Graphical representation of the aggregation model*

Let us illustrate the MYRIAD model previously constructed on the two KPIs "Q-score MaxCut" and "Q-score MaxClique" on two QPU – namely D-Wave 2000Q and D-wave Advantage.

| | Q-score MaxCut | Q-score MaxClique |
|---|---|---|
| D-Wave 2000Q | 70 | 70 |
| D-Wave Advantage | 140 | 110 |

The results of the evaluations are displayed in Figure 13 on evaluation of the two QPUs.

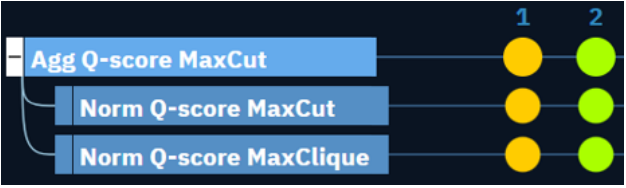

*Figure 13: Evaluation of the two QPUs. Option 1 is D-Wave 2000Q, and option 2 is D-Wave Advantage*

As an explanation of the evaluations, we wish to show the contribution of each node in the model. The most relevant explanation is contrastive, that is comparative with respect to some reference situation. The explanation then takes the form of computing the level to which each node in the MCDA model contributes in the comparison of the alternative with the reference alternative [34]. Two reference alternatives are classically considered. The first one is called "worse" and corresponds to the worst possible alternative on all KPIs. With this alternative, we explain in which respect the solution is preferred to the worst possible case. This explanation highlights the assets of the solution. Figure 14 shows the explanation of the comparison of D-Wave Advantage in comparison to the best possible alternative. For each node (criterion or aggregation) in the hierarchy, the size of the bar corresponds to the level of contribution of this node in the overall comparison of the D-Wave Advantage in comparison to the best possible alternative. It follows that "Q-score MaxCut" has an impact of 56% whereas "Q-score MaxClique" has an impact of 44%.

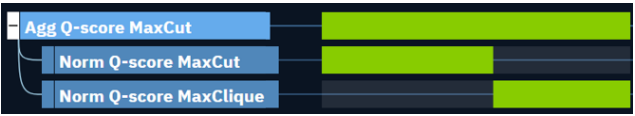

*Figure 14: Explanation of D-Wave Advantage in comparison to the best possible alternative*

## VI. FAST-TRACK QUANTUM COMPUTING BENCHMARKING INITIATIVE: PAVING THE WAY FOR COLLABORATIVE PROGRESS

The Fast-Track initiative emphasizes a collective effort, driven by collaboration among industry leaders and institutions. Its primary objective is to conduct a comprehensive measurement campaign on existing Quantum Processing Units (QPUs), utilizing the established metrics of Q-score [5]. This marks a pivotal juncture in the evaluation and benchmarking of quantum computing.

### A. Fast-Track broad collaboration

At the forefront of the Fast-Track initiative is a collaborative consortium strategically reaching out to the French quantum ecosystem, followed by European counterparts and concluding with global industry leaders. This inclusive group includes IQM, Pasqal, Quandela and Tu-Delft (TNO) alongside notable collaborators such as AQT, Quantinuum, IonQ, and QuEra. Some partners, such as the startups A&B, C12, Quobly, QB and Xanadu, express willingness to engage with us later with improved hardware or on other metrics. Each partner brings unique perspectives and expertise to drive significant advancements in quantum computing benchmarks. For instance, QuEra explores quantum algorithms and engages in discussions to contribute meaningfully. Xanadu explores collaboration possibilities, while A&B expresses eagerness in defining benchmarks that align with Fast-Track's broader goals. IonQ, despite initial reservations, explores potential involvement in BACQ, and AQT, with notable quantum volume achievements, deliberates on digital and analog approaches for Q-score MaxCut. Additionally, the initiative has collaborated with several partners on the formulation of MaxCut and its integration into specific qubit technologies, further enriching the collaborative and dynamic nature of the Fast-Track initiative and contributing to a richer quantum computing benchmarking landscape.

Collaborations with IQM, Pasqal, Quandela and TNO have progressed significantly, showcasing tangible advancements in Q-score MaxCut testing. IQM focuses on the successful resolution of Q-score MaxCut, addressing challenges to refine benchmarking accuracy. Pasqal, leveraging their emulator capabilities and actively engaging in Q-score testing on their quantum emulator, significantly contributes to benchmark evaluations [46]. Quandela's emulator-based QPU substantially enhances the testing environment, demonstrating a strong commitment to achieving benchmarking milestones. TNO's benchmarking initiative aligns cohesively with Fast-Track's goals, including a comprehensive exploration of quantum capabilities, such as testing Q-scores on D-Wave quantum devices [32], fostering robust collaboration.

### B. Fast-Track open collaboration

The Fast-Track initiative extends an open invitation to the wider quantum computing community, welcoming new partners to contribute to this transformative initiative. While some collaborations have matured with committed partners, the initiative remains inclusive and dynamic, actively seeking additional contributors who share a passion for advancing quantum computing benchmarks.

### C. Fast-Track conclusion

The Fast-Track Quantum Computing Benchmarking Initiative stands as evidence of collaborative innovation in the quantum computing arena. With advanced partnerships propelling the initiative forward and an open invitation for new contributors, Fast-Track is poised to deliver results that will significantly shape the future of quantum computing benchmarks. As the quantum computing community unites, the initiative marks a new era of exploration and collaboration, propelling us closer to unlocking the full potential of quantum computing.

ACKNOWLEDGMENT

As part of the MetriQs-France program, this work is supported by France 2030 under the French National Research Agency grant number ANR-22-QMET-0002.

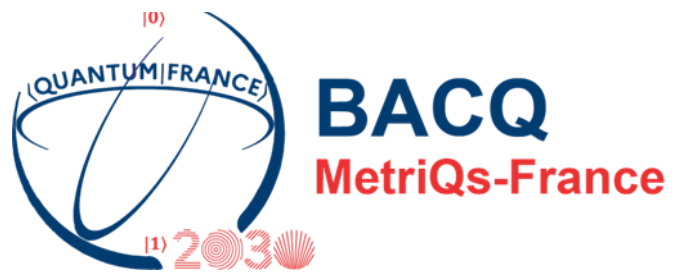

*Figure 15: BACQ logo*